\documentclass[11pt,a4paper]{article}
\pdfoutput=1

\usepackage{jcappub}
\usepackage{amssymb,amsmath,graphicx,latexsym}
\usepackage{bm}
\usepackage{epsfig}
\usepackage{epstopdf}

\usepackage{dcolumn}

\usepackage{a4wide}

\usepackage{amsthm}

\usepackage{amsfonts}

\usepackage{multicol}

\usepackage{multirow}
\def\Planck{\textit{Planck}}
\def\CCCP{\textit{CCCP}}

\title{Cosmological parameter estimation from CMB and X-ray clusters after Planck}

\author{Jian-Wei Hu,\footnote{Email: jwhu@itp.ac.cn}}
\author{Rong-Gen Cai,\footnote{Email: cairg@itp.ac.cn}}
\author{Zong-Kuan Guo,\footnote{Email: guozk@itp.ac.cn}}
\author{Bin Hu\footnote{Email: hu@lorentz.leidenuniv.nl}}
\affiliation{$^1$ $^2$ $^3$State Key Laboratory of Theoretical Physics, Institute of Theoretical Physics, Chinese Academy of Sciences, P.O. Box 2735, Beijing 100190, China}
\affiliation{$^4$ Instituut-Lorentz for Theoretical Physics,
Universiteit Leiden, 2333 CA Leiden, The Netherlands}

\date{\today}

\abstract{We update the cosmological parameter estimation for three non-vanilla models
by a joint analysis of \CCCP\ X-ray cluster, the newly released \Planck\ CMB data
as well as some external data sets, such as baryon acoustic oscillation measurements
from the 6dFGS, SDSS DR7 and BOSS DR9 surveys, and Hubble Space Telescope $H_0$ measurement.
First of all, we find that X-ray cluster data sets strongly favor a non-zero summed neutrino mass
at more than 3$\sigma$ confidence level in these non-vanilla models.
And then, we reveal some tensions between X-ray cluster and {\it Planck} data in some cosmological parameters.
For the matter power spectrum amplitude 
$\sigma_8$, X-ray cluster data favor a lower value compared with {\it Planck}.
Because of the strong $\sigma_8-\sum m_{\nu}$ degeneracy, this tension could beyond 2$\sigma$ confidence level
when the summed neutrino mass $\sum m_{\nu}$ is allowed to vary.
For the CMB lensing amplitude $A_L$, the addition of X-ray cluster data results in a 3$\sigma$
deviation from the vanilla model.
Furthermore, {\it Planck}+X-ray data prefer a large Hubble constant and
phantom-like dark energy equation of state,
which are in $2\sigma$ tension with those from WMAP7+X-ray data.
Finally, we find that these tensions/descrepencies could be relaxed in some sense
by adding a $9\%$ systematic shift in the cluster mass functions.
}

\arxivnumber{0000.0000}

\begin{document}
\maketitle

\section{Introduction}
Recently, the {\it Planck} Collaboration
publicly released the initial cosmology products \cite{Ade:2013ktc} based on the first
15.5 months of {\it Planck} operations.
Their scientific results strongly support the standard
6-parameter $\Lambda$CDM model, hereafter namely the vanilla model.
And the corresponding parameter constraints are greatly improved, including a
highly significant deviation from scale invariance of the primordial power spectrum of curvature perturbations.
However, some based cosmological parameter values and others derived from them
are significantly different from those previously determined,
such as present Hubble parameter $H_0$ \cite{Planck_16},
the lensing amplitude $A_{\rm L}$ \cite{Planck_16}, {\it etc.}
Among these parameter estimation tensions, the most controversial one is about $H_0$ value.
On the one hand, {\it Planck} results are discrepant
with recent direct measurements from Hubble Space Telescope (HST) Key Project \cite{Freedman:2000cf}
and Type Ia supernovae observations via the magnitude-redshift relation, such as the Union2.1 compilation \cite{Suzuki:2011hu}.
On the other hand, they are in excellent agreement with geometrical constraints from several
baryon acoustic oscillation (BAO) surveys \cite{Beutler:2011hx,Percival:2009xn,Anderson:2012sa}.
Beside that, very recently the authors of \cite{Spergel:2013rxa} re-analyze the {\it Planck} primary CMB data
and find that the 217 GHz $\times$ 217 GHz detector set spectrum used in the {\it Planck}
analysis is responsible for some of this tension.
In order to reveal or reconcile the tensions between low redshift geometric and {\it Planck}
measurements, many efforts have been done
\cite{Cheng:2013csa,Marchini:2013oya,Hu:2013aqa,Wyman:2013lza,Xia:2013dea,Hamann:2013iba,Battye:2013xqa,Li:2013dwy,Cai:2013owa}.

Beside CMB observations, Large Scale Structure (LSS) surveys on various scales, such as galaxies and clusters of galaxies,
could also provide us lots of cosmological information.
A better understanding of the sturcture of our universe asks for the agreement between theoretical predictions and observations on various
spatial and temporal scales. The cosmological information encoded in the CMB map is mainly on the
spatially large scales and at temporally very deep redshift. As complementary observations, the distribution of
LSS tells us the structure formation laws due to the
instability of gravity on relatively small scales and at low redshifts.
As the most massive virialised structures in the universe, clusters of galaxies are perfect
probes of the matter distribution on large scales. Within the framework of dark matter structure formation
scenario, baryonic matter traces the distribution of dark matter halo. When the baryonic gas falls into the
gravitational potential wells, it could heat up to $10^7$K so that X-rays will be emitted. Via this mechanism
the galaxy clusters could be identificated through their X-ray flux.
\textit{Chandra} Cluster Cosmology Project (\CCCP) \cite{Vikhlinin:2005mp,Kotov:2005uv,Kravtsov:2006db,Nagai:2006sz}
utilizes X-rays indicator to observe the galaxy clusters that has been a catalog detected in a new
Rontgensatellite (Rosat) PSPC surveys \cite{Burenin:2006td}
covering 400 square degrees sky area. Thanks to the high resolution of the \textit{Chandra} X-rays observator,
high-quality X-ray data of the resulting samples
upto redshift $z=0.9$ are obtained, which can be used to determine the galaxy cluster mass function and hence to estimate the cosmological parameters.

The detection of solar and atmospheric neutrino oscillations indicates
that neutrinos are massive, but cannot provide absolute masses for neutrinos.
Cosmological observations can provide significantly strong constraints on the
summed neutrino mass through the cosmological effects of massive neutrinos.
Neutrino masses affect the CMB power spectrum mainly through the early integrated
Sachs-Wolfe effect, the BAO by changing the late-time expansion rate of the universe,
and the abundance of galaxy clusters by smearing out a fraction of the mass
over the neutrino free streaming scale~\cite{Hou:2012xq}. The Planck team actually adopts a normal hierarchy for neutrino masses
with $\sum m_\nu=0.06$ eV as their baseline model
and finds a significant discrepancy between the Planck data and the abundance of galaxy clusters~\cite{Planck_16}.
This suggests that the tension is relaxed by increasing the summed neutrino mass
because their free streaming reduces the amount of small scale clustering today.
Therefore, in our analysis the summed neutrino mass is always allowed to be free.
In this paper we focus on the cosmological parameter estimation for three non-vanilla models
by using the {\it Planck}+WP+BAO+HST data in combination with {\it CCCP} X-ray clusters.

The rest parts of this paper are organized
as follows. In Sec. 2 we will briefly describe the data sets and methodology.
In Sec. 3 we will present our results in three non-vanilla models and reveal some tensions between
X-ray cluster and {\it Planck} data. Finally we arrive at our conclusions in Sec. 4.

\begin{table}[htb]
\scriptsize
\begin{center}
\caption{List of cosmological parameters}
\label{tab:parameters}
\begin{tabular}{|c|c|c|c|}
\hline
\hline

Parameter                    & Range                            & Baseline        & Definition                            \\  \hline

$\Omega_{\mathrm{b}}h^2$     & $[0.005, 0.1]$                   & --              & Baryon density today     \\
$\Omega_{\mathrm{c}} h^2$    & $[0.001, 0.99]$                  &  --             & Cold dark matter density today\   \\
$100\theta_{\mathrm{MC}}$    & $[0.5, 10.0]$                    & --              & Sound horizon parameter(CosmoMC)   \\
$\tau$                       & $[0.01, 0.8]$                    & --              & Thomson scattering optical depth of reionization   \\
$n_\mathrm{s}$               & $[0.9, 1.1]$                     & --              & Scalar spectrum power-law index      \\
$\ln(10^{10} A_\mathrm{s})$  & $[2.7, 4.0]$                     & --              & Amplitude of primordial curvature perturbations  \\
$\Sigma m_\nu\,[\mathrm{eV}]$& $[0, 5]$                         & --              & The sum of neutrino masses in eV        \\ \hline
\hline
$N_{\mathrm{eff}}$           & $[0.05, 10.0]$                   & 3.046           & Effective number of neutrino-like relativistic degrees of freedom  \\ \hline
$w$                          & $[-3.0, -0.3]$                   & $-1$              & Dark energy equation of state      \\ \hline
$\Omega_K$                   & $[-0.3, 0.3]$                    & 0               & Curvature parameter today \\ \hline
$A_L$                        & $[0,  10]$                       & 1               & Normalized lensing spectrum amplitude\\ \hline
\end{tabular}
\end{center}
\end{table}

\section{Data and methodology}

The total \Planck\ CMB temperature power-spectrum likelihood is divided into low-$l$ ($l<50$) and high-$l$ ($l\geq 50$) parts.
This is because the central limit theorem ensures that the distribution of CMB angular power spectrum $C_l$ in the high-$l$ regime
can be well approximated by Gaussian statistics. However, for the low-$l$ part the $C_l$ distribution is non-Gaussian.
For this reason the \Planck\ team adopts two different methodologies to build the likelihood. In detail,
for the low-$l$ part, the likelihood exploits all \Planck\ frequency channels
from $30$ to $353$ GHz, separating the cosmological CMB signal from diffuse Galactic foregrounds through a
physically motivated Bayesian component separation technique.
For the high-$l$ part, the \Planck\ team employs a correlated Gaussian likelihood approximation,
based on a fine-grained set of angular cross-spectra derived from multiple detector combination between the $100$, $143$, and $217$ GHz
frequency channels, marginalizing over power-spectrum foreground templates. In order to break the
well-known parameter degeneracy between the reionization optical depth $\tau$ and the scalar spectral index $n_s$, the
\Planck\ team adopts the low-$l$ WMAP polarization likelihood (WP).

As stated above in this paper we are also interested in the X-ray cluster
data \cite{Vikhlinin:2008cd,Vikhlinin:2008ym,Burenin:2012uy}.
The \CCCP\ project measures cluster mass function
by using a high-redshift ($0.4<z<0.9$) subsample of the 400 square degree survey, 37 objects,
and low-redshift ($z<0.2$) subsample of the all-sky survey, 49 brightest clusters.
The methodologies of likelihood construction follow the standard derivation of
the Poisson distribution of cluster mass \cite{Cash:1979vz}.
The likelihood function implicitly depends on the cosmological parameters through the model
of cluster mass function (reflecting the growth, normalization, and shape of the density perturbation power spectrum),
through the cosmological volume-redshift relation which determines the survey volume, and through the distance-redshift
as well as the masses-temperature relation.
The details of likelihood construction and systematic uncertainty control are mentioned in \cite{Vikhlinin:2008cd}.
Furthermore, in order to break the parameter degeneracies we also use some other external data sets,
including BAO measurements from the 6dFGS \cite{Beutler:2011hx},
SDSS DR7 \cite{Percival:2009xn} and BOSS DR9 \cite{Anderson:2012sa} surveys,
and HST Key project \cite{Freedman:2000cf} $H_0$ measurement.

Contrary to the {\it Planck} constraints on the summed neutrino mass,
\cite{Wyman:2013lza} found that adding X-ray cluster could gives the
non-zero detection of the active or sterile neutrino mass ($\sum m_{\nu}$ or $m_s$) with great statistical significance
for various 8-parameter models, including the active/sterile neutrino mass as well as effective neutrino number $N_{\rm eff}$.
Moreover, adding X-ray data set could also lead to
a significant deviation in $\sigma_8$, namely the matter power spectrum amplitude on the 8$h^{-1}$Mpc scale,
from the {\it Planck} result~\cite{Planck_16}.
In details, without X-ray cluster, the {\it Planck} result favors a larger value of $\sigma_8$,
while with them the joint analysis give a lower value.
Therefore, in this paper we explore the tension between {\it Planck} and {\it CCCP} X-ray cluster data sets with
several 8-parameter models, including effective neutrino number $N_{\rm eff}$, constant dark energy equation of state $w$,
present spatial curvature $\Omega_K$ and lensing amplitude $A_{L}$.
Particularly,
we here focus on the neutrino mass constraint, so that in our baseline model
the summed neutrino mass is always allowed to vary freely.
We restrict ourselves to one-parameter extensions to the baseline model of $\Lambda{\rm CDM}+\sum m_\nu$, as listed
in Tab.\ref{tab:parameters}.

We compute the CMB angular and matter power spectra by using the public Einstein-Boltzmann solver \emph{CAMB}~\cite{Lewis:1999bs}
and explore the cosmological parameter space with a Markov Chain Monte Carlo sampler, namely \texttt{CosmoMC} \cite{Lewis:2002ah}.
For {\it Planck} we use the {\it Planck} Likelihood Code (PLC/clik) \cite{Planck:2013kta} which are available at the Planck Legacy Archive \cite{PLA},
and for {\it CCCP} our analysis is based on the likelihood grids presented in \cite{Vikhlinin:2008cd},
which can be download from the website~\cite{400d}.

\section{Results}
In this section we will dig the information hidden in the {\it CCCP} X-ray cluster data \cite{Vikhlinin:2008cd,Vikhlinin:2008ym,Burenin:2012uy}.
Hereafter, we denote $CL_{\rm X-ray}$ for this data.
We will first investigate the
constraints on the summed neutrino mass from the {\it Planck}+WP+BAO+HST data and $CL_{\rm X-ray}$. Hereafter, we dub {\it Planck}+WP+BAO+HST as PWBH.
Then, we will turn to reveal some tensions in some cosmological
parameters between these two data sets.

\subsection{$\sum m_{\nu}$ results}
First, let us study the summed neutrino mass $\sum m_{\nu}$.
The solar and atmospheric oscillation observations have already set a lower bound ($\sum m_{\nu}\geq 0.06$~eV)
on the summed mass for the standard three neutrino species. Beside the local observations, we could also
obtain neutrino mass information via the indirect measurements on the cosmological scales.
Generally speaking, there are mainly two ways. One is through the secondary CMB anisotropies generated in the
deep matter dominated epoch, such as weak lensig effect. However, these anisotropies are so small compared with
the primordial signal that the current CMB experiments could only give a very loose upper bound, such as $\sum m_{\nu}<0.66$ eV \cite{Planck_16}
from {\it Planck} \cite{Ade:2013ktc}+ACT \cite{Das:2013zf}+SPT \cite{Keisler:2011aw,Story:2012wx,Reichardt:2011yv}.
The other method is to utilize the large scale structure tracers, such as matter power spectrum, selected cluster counting
and cosmic shear, {\it etc.} Compared with the bounds obtained by CMB observations, these tomographic measurements could set a relatively
stringent constraint. However, due to the contaminations from systematic noise and theoretical non-linearity {\it etc.},
the resulting constraint varies a lot among different projects.
For example, $CL_{\rm X-ray}$ \cite{Wyman:2013lza} and selected Sunyaev-Zel'dovich galaxy
cluster counts from {\it Planck} \cite{Ade:2013lmv,Battye:2013xqa} and SPT \cite{Hou:2012xq}
report the non-zero detection of summed neutrino mass with quite significant evidence, while
other galaxy surveys, such as WiggleZ \cite{Riemer-Sorensen:2013jsa} could improve the upper bound a lot but do not find any
significant deviation from zero.


\begin{table}[tmb]
\footnotesize
\begin{center}
\caption{\Planck+WP+BAO+HST+$CL_{\rm X-ray}$ results}
\begin{tabular}{|c|cc|cc|cc|}
\hline
\hline

\multirow{2}{*}{Model}       & \multicolumn{2}{c|}{$\Lambda$CDM+$\Sigma{m_\nu}$+$N_{\rm eff}$} & \multicolumn{2}{c|}{$w$CDM+$\Sigma{m_\nu}$} & \multicolumn{2}{c|}{$\Lambda$CDM+$\Sigma{m_\nu}$+$\Omega_{K}$}\\\cline{2-7}

                             &best fit    & 68\% limits         & best fit  & 68\% limits            & best fit         & 68\% limits          \\ \hline

100$\Omega_{\mathrm{b}}h^2$  & 2.285      & 2.274$\pm$0.027     & 2.205     & 2.204$\pm$0.026        & 2.224            & 2.203$\pm$0.031      \\
$\Omega_{\mathrm{c}} h^2$    & 0.1242     & 0.1227$\pm$0.0044   & 0.1181    & 0.1176$\pm$0.0016      & 0.1172           & 0.1179$\pm$0.0026    \\
$100\theta_{\mathrm{MC}}$    & 1.04085    & 1.04092$\pm$0.00070 & 1.04139   & 1.04125$\pm$0.00057    & 1.04190          & 1.04125$\pm$0.00065  \\
$\tau$                       & 0.095      & 0.097$\pm$0.015     & 0.086     & 0.090$\pm$0.013        & 0.089            & 0.089$\pm$0.013      \\
$n_\mathrm{s}$               & 0.9925     & 0.9929$\pm$0.0098   & 0.9595    & 0.9621$\pm$0.0059      & 0.9654           & 0.9624$\pm$0.0077    \\
$\ln(10^{10} A_\mathrm{s})$  & 3.107      & 3.109$\pm$0.031     & 3.075     & 3.081$\pm$0.025        & 3.083            & 3.080$\pm$0.026      \\
$\Sigma m_\nu\,[\mathrm{eV}]$& 0.47       & 0.46$\pm$0.12       & 0.56      & 0.55$\pm$0.10          & 0.41             & 0.45$\pm$0.12        \\ \hline
$N_{\rm eff}$           & 3.80       & 3.704$\pm$0.29      & --        & --                     & --               & --                   \\
$w$                     & --         & --                  & $-1.39$     & $-1.39\pm$0.12         & --               & --                   \\
$\Omega_K$                   & --         & --                  & --        & --                     & 0.00695          & 0.00835$\pm$0.00411  \\ \hline
$\Omega_{\mathrm{m}}$       & 0.2989     & 0.3000$\pm$0.012     & 0.2693    & 0.2661$\pm$0.0139      & 0.3027           & 0.3081$\pm$0.0131    \\
$H_0$                       & 71.36      & 70.8$\pm$1.5         & 73.68     & 74.0$\pm$2.1           & 68.93            & 68.6$\pm$1.0         \\
$\sigma_8$                  & 0.7461     & 0.7477$\pm$0.0151    & 0.7902    & 0.7937$\pm$0.0208      & 0.7434           & 0.7370$\pm$0.0164    \\ \hline
$\chi_{\rm min}^2/2$            & \multicolumn{2}{c|}{4911.581}     & \multicolumn{2}{c|}{4908.152}      & \multicolumn{2}{c|}{4912.452} \\ \cline{1-7}
\end{tabular}
\label{Tab:mnu}
\end{center}
\end{table}

\begin{figure*}[tmb]
\begin{center}
\includegraphics[scale=0.4]{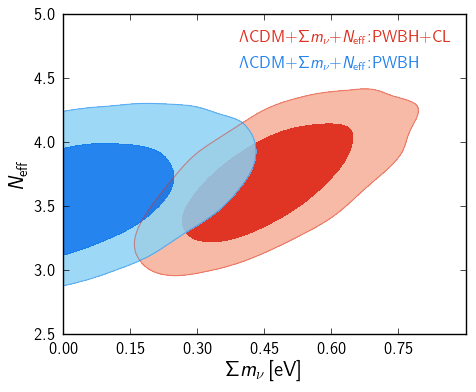}\includegraphics[scale=0.4]{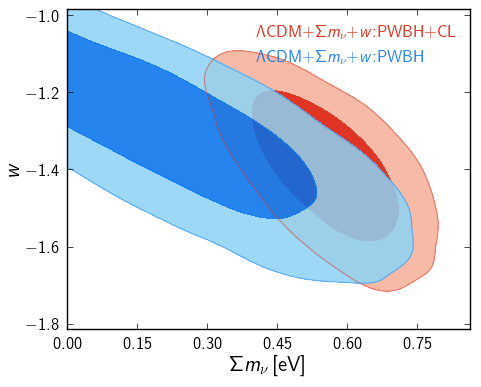}\includegraphics[scale=0.4]{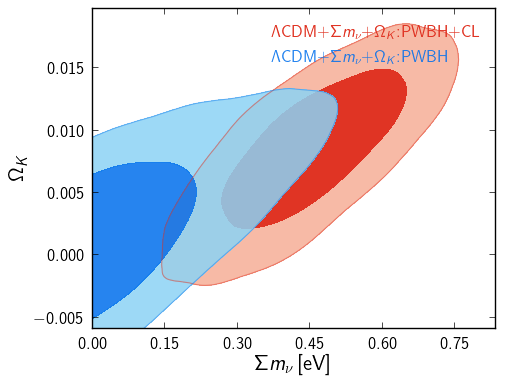}
\caption{\emph{Left}: Likelihood contours (68\% CL and 95\% CL) in the $\sum m_{\nu}$--$N_{\mathrm{eff}}$ plane for
the \textit{Planck}+WP+BAO+$H_{0}$+$CL_{\rm {X-ray}}$(red) and \textit{Planck}+WP+BAO+$H_{0}$ (blue) data combinations.
\emph{Middle}: $\sum m_{\nu}$--$w$ likelihood contours.
\emph{Right}:  $\sum m_{\nu}$--$\Omega_{K}$ likelihood contours.}
\label{mnu}
\end{center}
\end{figure*}

Given the above facts, in what follows we will use $CL_{\rm X-ray}$ data to do the joint
analysis of the summed neutrino mass $\sum m_{\nu}$ with several related parameters,
such as effective neutrino number $N_{\rm eff}$, dark energy equation
of state $w$ as well as spatial curvature $\Omega_K$. The results with or without $CL_{\rm X-ray}$
are listed in Tab.\ref{Tab:mnu} or Tab.\ref{Tab:PWBH}. And the corresponding 2D likelihood contours are shown
in Fig.\ref{mnu}.

\begin{table}[tmb]
\footnotesize
\begin{center}
\caption{\Planck+WP+BAO+HST results}
\begin{tabular}{|c|cc|cc|cc|}
\hline
\hline

\multirow{2}{*}{Model}       & \multicolumn{2}{c|}{$\Lambda$CDM+$\Sigma{m_\nu}$+$N_{\rm eff}$} & \multicolumn{2}{c|}{$w$CDM+$\Sigma{m_\nu}$} & \multicolumn{2}{c|}{$\Lambda$CDM+$\Sigma{m_\nu}$+$\Omega_{K}$}\\\cline{2-7}

                             &best fit    & 68\% limits         & best fit  & 68\% limits            & best fit         & 68\% limits          \\ \hline

100$\Omega_{\mathrm{b}}h^2$  & 2.225      & 2.249$\pm$0.027     & 2.183     & 2.193$\pm$0.026        & 2.214            & 2.210$\pm$0.031     \\
$\Omega_{\mathrm{c}} h^2$    & 0.1241     & 0.1275$\pm$0.0048   & 0.1225    & 0.1209$\pm$0.0022      & 0.1188           & 0.1192$\pm$0.0028   \\
$100\theta_{\mathrm{MC}}$    & 1.04090    & 1.04060$\pm$0.00071 & 1.04050   & 1.04105$\pm$0.00059    & 1.04137          & 1.04138$\pm$0.00066   \\
$\tau$                       & 0.091      & 0.097$\pm$0.014     & 0.086     & 0.089$\pm$0.013        & 0.092            & 0.091$\pm$0.013   \\
$n_\mathrm{s}$               & 0.9743     & 0.9830$\pm$0.0097   & 0.9552    & 0.9571$\pm$0.0065      & 0.9628           & 0.9619$\pm$0.0077     \\
$\ln(10^{10} A_\mathrm{s})$  & 3.102      & 3.122$\pm$0.030     & 3.087     & 3.090$\pm$0.025        & 3.092            & 3.091$\pm$0.025      \\
$\Sigma m_\nu\,[\mathrm{eV}]$& 0.039      & $<0.34$ (95\% CL)& 0.17  & $<0.61$ (95\% CL)  & 0.018 & $<0.38$ (95\% CL)        \\ \hline
$N_{\mathrm{eff}}$           & 3.44       & 3.661$\pm$0.27      & --        & --                     & --               & --                   \\
$w$                     & --         & --                  & $-1.27$     & $-1.33\pm$0.15         & --               & --                   \\
$\Omega_K$                   &--          &--                   & --        & --                     & 0.0013           & 0.00318$\pm$0.00415 \\ \hline
$\Omega_{\mathrm{m}}$       & 0.2948     & 0.2986$\pm$0.011     & 0.2778    & 0.2693$\pm$0.0150      & 0.2978           & 0.3031$\pm$0.0112   \\
$H_0$                       & 70.57      & 71.2$\pm$1.5         & 72.53     & 73.7$\pm$2.3           & 68.84            & 68.6$\pm$1.0            \\
$\sigma_8$                  & 0.8463     & 0.8403$\pm$0.0271    & 0.8828    & 0.8612$\pm$0.0388      & 0.8384           & 0.8164$\pm$0.0285   \\ \hline
$\chi_{\rm min}^2/2$            & \multicolumn{2}{c|}{4903.787}     & \multicolumn{2}{c|}{4903.607}      & \multicolumn{2}{c|}{4905.298} \\ \cline{1-7}
\end{tabular}
\label{Tab:PWBH}
\end{center}
\end{table}

We summarise our main results in what follows.
First of all, from Tab.\ref{Tab:mnu} one could find a more than $3\sigma$ detection of the summed neutrino mass
for these three non-vanilla models when $CL_{\rm {X-ray}}$ data are taken into account
\begin{eqnarray}
 \sum m_{\nu} &=&0.46\pm0.12 \ (68\%~;~+N_{\rm eff}:{\rm PWBH}+CL)\;,\\
 \sum m_{\nu} &=&0.55\pm0.10 \ (68\%~;~+w:{\rm PWBH}+CL) \;,\\
 \sum m_{\nu} &=&0.45\pm0.12 \ (68\%~;~+\Omega_K:{\rm PWBH}+CL) \;.
\end{eqnarray}
Moreover, as pointed out in \cite{Burenin:2012uy} there exists a systematical error in hydrostatic mass measurements
$\delta M/M \simeq 0.09$ in $CL_{\rm X-ray}$ data sets. So, we take this mass function correction into account and
list the corresponding results in Tab.\ref{Tab:9mass}.
Thanks to this correction, the mean value of summed neutrino mass get reduced
\begin{equation}
    \sum m_{\nu} = 0.39\pm 0.09  \     (68\%~;~+w:{\rm PWBH}+CL+9\% {\rm Mass}).
\end{equation}

\begin{table}[htb]
\footnotesize
\begin{center}
\caption{\Planck+WP+BAO+HST+$CL_{\rm {X-RAY}}$+9\% Mass result }
\begin{tabular}{|c|cc|cc|}
\hline
\hline

\multirow{2}{*}{Model}       & \multicolumn{2}{c|}{$w$CDM+$\Sigma{m_\nu}$}  & \multicolumn{2}{c|}{$w$CDM+$\Sigma{m_\nu}$+$A_L$}
\\ \cline{2-5}

                             &best fit    & 68\% limits            & best fit  & 68\% limits
                             \\ \hline

100$\Omega_{\mathrm{b}}h^2$  & 2.213      & 2.214$\pm$0.024        & 2.226     & 2.243$\pm$0.027
\\
$\Omega_{\mathrm{c}} h^2$    & 0.1178     & 0.1170$\pm$0.0013      & 0.1182    & 0.1166$\pm$0.0014
\\
$100\theta$                  & 1.04175    & 1.04143$\pm$0.00056    & 1.04129   & 1.04165$\pm$0.00059
\\
$\tau$                       & 0.086      & 0.089$\pm$0.013        & 0.081     & 0.087$\pm$0.013
\\
$n_\mathrm{s}$               & 0.9656     & 0.9648$\pm$0.0054     & 0.9665    & 0.9685$\pm$0.0057
\\
$\ln(10^{10} A_\mathrm{s})$  & 3.077      & 3.080$\pm$0.025        & 3.069     & 3.076$\pm$0.024
\\
$\Sigma m_\nu\,[\mathrm{eV}]$& 0.40       & 0.39$\pm$0.09          & 0.42      & 0.38$\pm$0.09
\\ \hline
$A_{L}$                      & --         & --                     & 1.22      & 1.28$\pm$0.10         \\
$w$                     & $-1.28$      & $-1.23\pm$0.057        & $-1.29$     & $-1.20\pm$0.07         \\ \hline
$\Omega_{\mathrm{m}}$        & 0.2687     & 0.2766$\pm$0.0113      & 0.2726    & 0.2780$\pm$0.0118
\\
$H_0$                        & 73.26      & 72.0$\pm$1.3           & 72.93     & 71.8$\pm$1.5
\\
$\sigma_8$                   & 0.8109     & 0.7977$\pm$0.0170      & 0.8031    & 0.7639$\pm$0.0198
\\ \hline
$\chi_{\rm min}^2/2$            & \multicolumn{2}{c|}{4907.444}     & \multicolumn{2}{c|}{4903.368}       \\ \cline{1-5}
\end{tabular}
\label{Tab:9mass}
\end{center}
\end{table}

\subsection{$A_L$ and $\sigma_8$ results}

\begin{table}[tmb]
\footnotesize
\begin{center}
\caption{CMB lensing amplitude $A_L$ results}
\begin{tabular}{|c|cc|cc|cc|}
\hline
\hline

\multirow{2}{*}{Model}       & \multicolumn{2}{c|}{$\Lambda$CDM+$\Sigma{m_\nu}$+$A_{L}$}   & \multicolumn{2}{c|}{$\Lambda$CDM+$A_{L}$} & \multicolumn{2}{c|}{$\Lambda$CDM+$A_L$ (without {\rm HST}+$CL_{\rm {X-ray}}$)}\\ \cline{2-7}

                             &best fit    & 68\% limits         & best fit  & 68\% limits            & best fit         & 68\% limits          \\ \hline

100$\Omega_{\mathrm{b}}h^2$  & 2.292      & 2.276$\pm$0.027     & 2.304     & 2.285$\pm$0.025       & 2.254           & 2.250$\pm$0.028    \\
$\Omega_{\mathrm{c}} h^2$    & 0.1132     & 0.1131$\pm$0.0011   & 0.1125    & 0.1124$\pm$0.0010     & 0.1167          & 0.1166$\pm$0.0017   \\
$100\theta_{\mathrm{MC}}$    & 1.04279    & 1.04224$\pm$0.00056 & 1.04202   & 1.04224$\pm$0.00056   & 1.04156         & 1.04186$\pm$0.00058 \\
$\tau$                       & 0.086      & 0.087$\pm$0.013     & 0.075     & 0.073$\pm$0.011       & 0.092           & 0.087$\pm$0.013    \\
$n_\mathrm{s}$               & 0.9763     & 0.9768$\pm$0.0053   & 0.9807    & 0.9784$\pm$0.0051     & 0.9703          & 0.9698$\pm$0.0059   \\
$\ln(10^{10} A_\mathrm{s})$  & 3.069      & 3.068$\pm$0.025     & 3.046     & 3.039$\pm$0.021       & 3.091           & 3.078$\pm$0.025      \\ \hline
$\Sigma m_\nu\,[\mathrm{eV}]$& 0.25       & 0.28$\pm$0.08       & --        & --                    & --              & --      \\
$A_{L}$                      & 1.42       & 1.36$\pm$0.10       & 1.44      & 1.37$\pm$0.11         & 1.24            & 1.22$\pm$0.10  \\\hline
$\Omega_{\mathrm{m}}$        & 0.2907     & 0.2942$\pm$0.0108   & 0.2696    & 0.2693$\pm$0.0057      & 0.2940          & 0.2929$\pm$0.0099    \\
$H_0$                        & 69.12      & 68.7$\pm$0.9        & 71.09     & 71.0$\pm$0.5           & 68.97           & 69.1$\pm$0.8      \\
$\sigma_8$                   & 0.7538     & 0.7550$\pm$0.0139   & 0.7894    & 0.7867$\pm$0.0070      & 0.8218          & 0.8162$\pm$0.0119    \\ \hline
$\chi_{\rm min}^2/2$            & \multicolumn{2}{c|}{4907.083}     & \multicolumn{2}{c|}{4910.408}      & \multicolumn{2}{c|}{4903.236} \\ \cline{1-7}
\end{tabular}
\label{Tab:AL}
\end{center}
\end{table}

\begin{figure}
\begin{center}
\includegraphics[width=10cm,height=10cm]{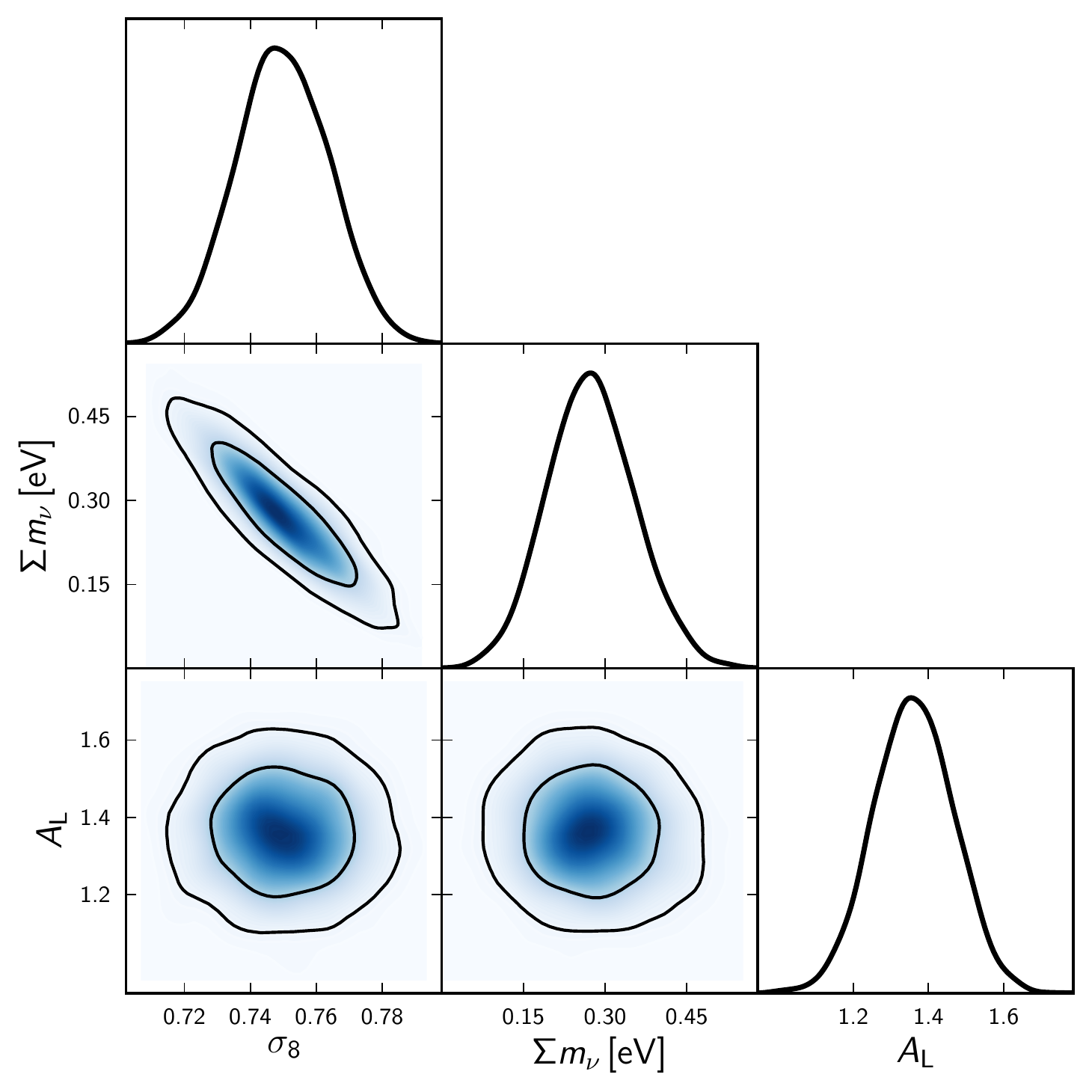}
\caption{Triangle likelihood contours of $\sigma_8$, $\sum m_{\nu}$ and $A_L$ with {\it Planck}+WP+BAO+HST+$CL_{\rm X-ray}$.}
\label{Fig:s8mnuAlens}
\end{center}
\end{figure}

Then we study models including $A_L$ and $\sigma_8$, in which the latter is considered as a
derived parameter from primordial curvature perturbation amplitude. Let us first investigate parameter degeneracies.
As shown in Fig.\ref{Fig:s8mnuAlens}, there exists only a tiny correlation between $\sum m_{\nu}$ and $A_L$,
so that the constraint on $A_L$ in $\Lambda$CDM+$\sum m_{\nu}$+$A_L$ and $\Lambda$CDM+$A_L$ models are very close (see Tab.\ref{Tab:AL}).
For example, using PWBH+$CL_{\rm X-ray}$ we could get
\begin{equation}
\label{AL_tension}
    A_L  =1.37\pm0.11 \    (68\%~;  \Lambda{\rm CDM}+A_L:{\rm PWBH}+CL)  \;.
\end{equation}
Furthermore, we list the results of $\Lambda$CDM+$A_L$ model without $CL_{\rm X-ray}$ in the third column of Tab.\ref{Tab:AL} for comparison,
which is well consistent with the {\it Planck} results \cite{Planck_16}.
It shows that without $CL_{\rm X-ray}$
\footnote{In order to compare with the {\it Planck} results \cite{Planck_16} we also remove HST data. As a background geometric
measurements, HST data sets should be nearly blind to dynamical structure formation information on
perturbation level. Hence, we should expect no significant change in $A_L$ value after removing HST data.}
\begin{equation}
\label{AL_tension2}
    A_L  =1.22\pm0.10 \      (68\%~;\Lambda{\rm CDM}+A_L: {\rm PWB})  \;.
\end{equation}
Comparing (\ref{AL_tension}) with (\ref{AL_tension2}) we could find that $CL_{\rm X-ray}$ leads the $A_L$ deviation from
unity, the value in vanilla model, even larger.
Moreover, in the second column of Tab.\ref{Tab:9mass}, we can see that
by adding $9\%$ mass correction, the tension in (\ref{AL_tension}) with the vanilla model could be mildly reconciled
\begin{equation}
    A_L  =1.28\pm0.10 \      (68\%~;w{\rm CDM}+\sum m_{\nu}+A_L:{\rm PWBH}+CL+9\%{\rm Mass})  \;.
\end{equation}

\begin{figure}
\begin{center}
\includegraphics[width=7cm,height=7cm]{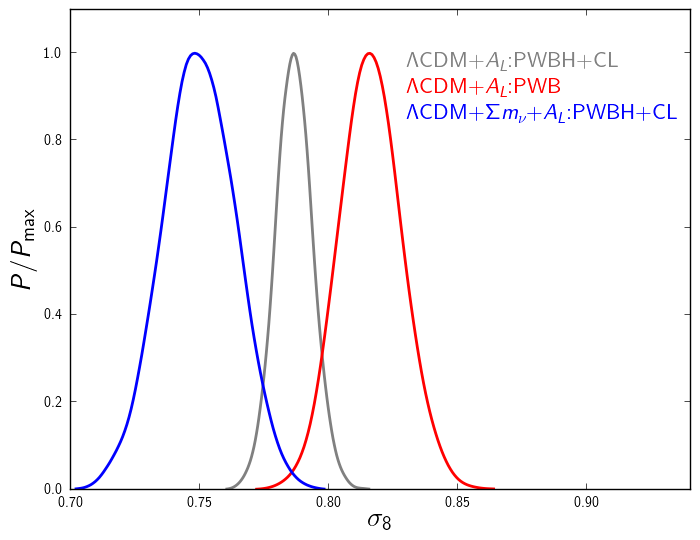}\includegraphics[width=7cm,height=7cm]{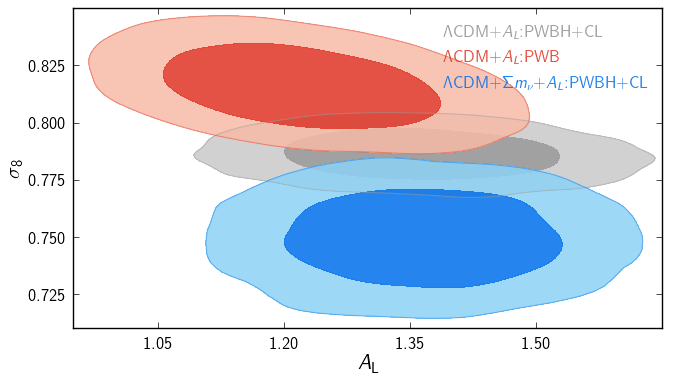}
\caption{{\it Left:} Marginalized likelihoods of $\sigma_8$.
{\it Right:} 2D likelihood contours between the lensing amplitude $A_{L}$ and rms amplitude of linear fluctuation $\sigma_{8}$.}
\label{Fig:AL}
\end{center}
\end{figure}

Then, we turn to the matter power spectrum amplitude $\sigma_8$.
As shown in Fig.\ref{Fig:s8mnuAlens}, there exists a significant anti-correlation between
$\sigma_8$ and $\sum m_{\nu}$.
This is because that the non-relativistic, massive, weakly-interacting neutrinos behave qualitatively as a species of
warm/hot dark matter, suppressing fluctuations on scales smaller than their thermal free-streaming length.
Consequently, this correlation will lead to a relatively low value of $\sigma_8$ when
$\sum m_{\nu}$ is allowed to vary (see Fig.\ref{Fig:AL} and Tab.\ref{Tab:AL}).
For example, by using data sets PWBH+$CL_{\rm {X-ray}}$ our results give
\begin{eqnarray}
    \sigma_8  &=&0.7894\pm0.0070 \    (68\%~;\ \Lambda{\rm CDM}+A_L:{\rm PWBH}+CL)  \;,\\
    \sigma_8  &=&0.7550\pm0.0139 \    (68\%~;\ \Lambda{\rm CDM}+\sum m_{\nu}+A_L:{\rm PWBH}+CL) \;.\label{s8vl}
\end{eqnarray}
In $6$ parameter $\Lambda$CDM model, {\it Planck} collaboration~\cite{Ade:2013lmv} gives $\sigma_8(\Omega_m/0.27)^{0.3}=0.784\pm0.027$ by using 
{\it Planck}+WP+BAO+BBN+$CL_{\rm {X-ray}}$, which is consistent with the results reported here.  
We notice that in Fig.\ref{Fig:AL}
the results with $CL_{\rm {X-ray}}$ (\ref{s8vl}) (blue curve) are in a $2\sigma$ tension with those in absence of $CL_{\rm {X-ray}}$ data (\ref{s8vh}) (red curve)
\begin{equation}
 \sigma_8  =0.8162\pm0.0119 \    (68\%~;\ \Lambda{\rm CDM}+A_L:{\rm PWB})  \;.\label{s8vh}
\end{equation}

Similar to $A_L$, with a $9\%$ cluster mass correction we could also reconcile the tension with {\it Planck}
\begin{equation}
    \sigma_8  =0.7639\pm0.0198 \    (68\%~;\ w{\rm CDM}+\sum m_{\nu}+A_L:{\rm PWBH}+CL+9\%{\rm Mass}) \;.
\end{equation}

\subsection{$H_0$ and $w$ results}
In this subsection, we study two background parameters, Hubble constant $H_0$
and Dark Energy (DE) Equation of State (EoS)  $w$.

\begin{figure*}[htb]
\begin{center}
\includegraphics[width=7cm,height=7cm]{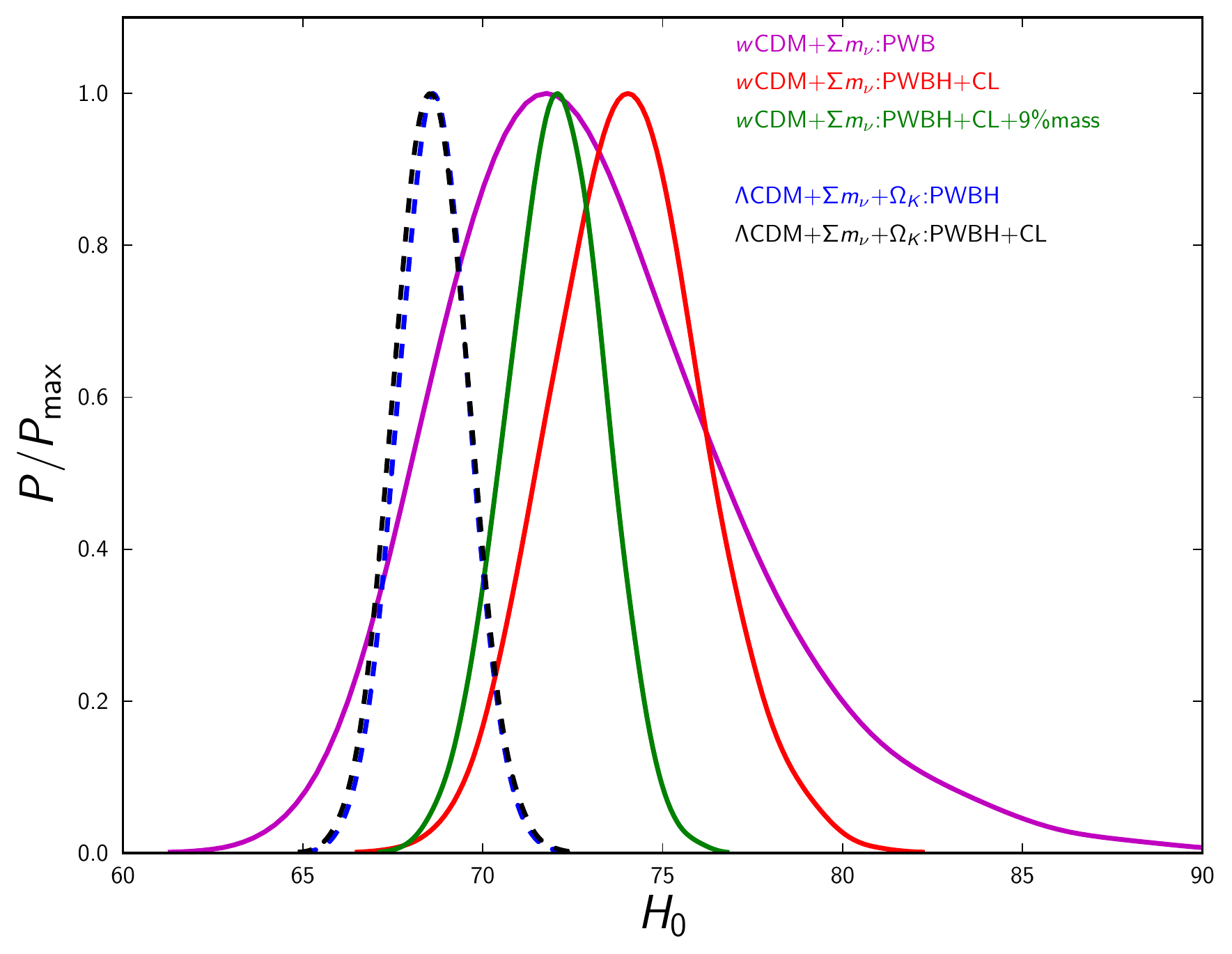}
\includegraphics[width=7cm,height=7cm]{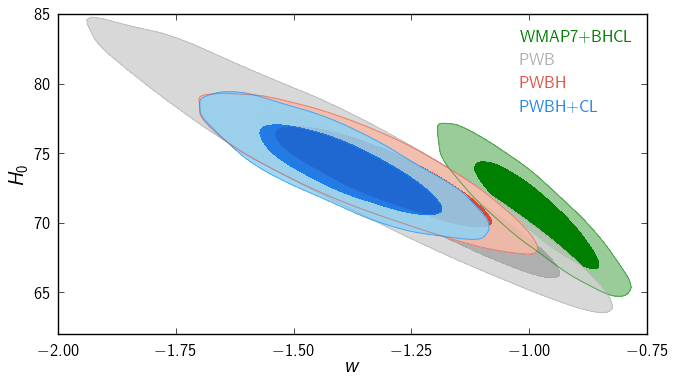}
\caption{ {\it Left:} marginalized likelihood of $H_0$ for
$w$CDM+$\Sigma{m_\nu}$ (solid) and $\Lambda$CDM+$\Sigma{m_\nu}$+$\Omega_{K}$ (dashed) model.
{\it Right:} 2D likelihood contours in the $H_0-w$ plane.}
\label{Fig:H0}
\end{center}
\end{figure*}

First of all, we can see from Tab.\ref{Tab:mnu} that without $9\%$ mass correction
to $CL_{\rm {X-ray}}$ data sets, a larger $H_0$ value is favored, e.g. for $w$CDM+$\sum m_{\nu}$ model:
\begin{equation}
 H_0=74.0\pm 2.1\; (68\%~;\;+w:{\rm PWBH}+CL)\;.
\end{equation}
However, after adding this corrections one could pull the mean value of $H_0$ a little bit back
for the same model
\begin{equation}
 H_0=72.0\pm 1.3\; (68\%~;\;+w:{\rm PWBH}+CL+9\%{\rm Mass})\;.
\end{equation}
Second, instead of the spatial curvature $\Omega_K$, once DE EoS $w$ is treated as a
free parameter, a larger $H_0$ arrives (see the left panel of Fig.\ref{Fig:H0}).
This is due to the $H_0-w$ correlation illustrated in the right panel of Fig.\ref{Fig:H0}.
Third, for a consistency check we also
include the results from joint analysis of WMAP7 and $CL_{\rm {X-ray}}$ data.
The right panel of Fig.\ref{Fig:H0} clearly shows that there exists a 2$\sigma$ tension in the parameter plane
between the {\it Planck}+WP+BAO+HST+$CL_{\rm {X-ray}}$ (blue)
and WMAP7+BAO+HST+$CL_{\rm {X-ray}}$ (green) data.

Beside the above $H_0$ descrepencies, we also notice that a phantom-like DE EoS
\cite{Caldwell:1999ew} is favored in our results
\begin{equation}
 w=-1.39\pm 0.12 (68\%~; +w:{\rm PWBH}+CL )\;.
\end{equation}
However, we should emphasize that this phantom-like value of DE EoS is not resulted in by including $CL_{\rm {X-ray}}$ data.
From the middel column of Tab.\ref{Tab:PWBH}, one can find that without X-ray cluster data
the DE EoS still deviates from $-1$ more than $2\sigma$ confidence level
\begin{equation}
    w = -1.33\pm0.15 \     (68\%~; +w:{\rm PWBH}).
\end{equation}
Furthermore, if we also remove HST data, i.e. just use \Planck+WP+BAO, there will be no significant deviation from $w=-1$ as
\begin{equation}
    w = -1.31\pm0.23 \     (68\%~;+w:{\rm PWB}).
\end{equation}
Because of the well-known $H_0-w$ correlation,
we argue that this result partially reflects the tension between HST and {\it Planck} on Hubble parameter $H_0$ as found in \cite{Planck_16}.
At last, the $9\%$ mass correction can also relax the tension in $w$ with $\Lambda$CDM model
\begin{equation}
    w =-1.23\pm 0.06  \     (68\%~;+w:{\rm PWBH}+CL+9\% {\rm Mass}).
\end{equation}

\section{Conclusions}
We have presented an updated estimation of cosmological parameters in three 8-parameter non-vanilla models using \Planck\, WMAP7, BAO, HST
and \CCCP\ X-ray data sets. First of all, we have found that X-ray cluster data sets strongly favor a non-zero summed neutrino mass
with more than 3$\sigma$ CL in these models, which is in quite good agreement with recent results
in the literature. The presence of massive neutrinos inhibits the growth of structures
below their thermal free-streaming scale during structure formation, leading to a reduced
value of $\sigma_8$, which could improve consistency with X-ray cluster data.
In addition, we have also revealed some tensions between X-ray cluster and {\it Planck} data
in the cosmological parameters, including the matter power spectrum
amplitude $\sigma_8$, lensing amplitude $A_L$, constant dark energy equation of state $w$ as well as Hubble parameter $H_0$.

For the matter power spectrum amplitude $\sigma_8$, X-ray cluster data favor a relatively low value compared with {\it Planck}.
Because of the $\sigma_8-\sum m_{\nu}$ degeneracy, this tension could beyond 2$\sigma$ CL
when the summed neutrino mass $\sum m_{\nu}$ is allowed to vary.
For the CMB lensing amplitude $A_L$, the addition of X-ray cluster data makes
its deviation from unity (vanilla model) even worse and
results in more than 3$\sigma$ descrepency.
Because of the correlation between $H_0$ and $w$, X-ray cluster data prefer a large Hubble constant and
quite negative dark energy equation of state. Furthermore, we have also found a
2$\sigma$ tension in the $H_0-w$ plane between the
{\it Planck}+WP+BAO+HST+$CL_{\rm {X-ray}}$ and WMAP7+BAO+HST+$CL_{\rm {X-ray}}$ data.
Finally, we should emphasize that these tensions/descrepencies could be reduced in some sense
by making a $9\%$ shift in the cluster mass functions.
The resolution of these intensions will likely require either the identification of
a currently-unknown systematic effect in at lease one of these data sets
or new physics.


\acknowledgments
This work is partially supported by the project of Knowledge
Innovation Program of Chinese Academy of Science,
NSFC under Grant No.11175225, No.11335012, and National Basic Research
Program of China under Grant No.2010CB832805 and No.2010CB833004. BH is supported
by the Dutch Foundation for Fundamental Research on Matter (FOM).


\end{document}